\documentclass[aps,pra,
 reprint,
a4paper,superscriptaddress,floatfix,amsmath,amssymb,amsfonts,noshowpacs,]{revtex4-1}
\usepackage{newtxtext,newtxmath}
\usepackage[utf8]{inputenx} 
\usepackage{graphicx}
\usepackage[dvipsnames]{xcolor}
\usepackage{soul} 
\usepackage[textwidth=17.5cm,textheight=23.5cm,verbose,pdftex]{geometry}
\usepackage{floatrow}
\usepackage{xspace}
\usepackage[pdftex]{hyperref}

\newcommand{\celsius}{\,$^{\circ}$C\xspace}
\newcommand{\mum}{\textmu{}m\xspace}
\newcommand{\grad}{$^{\circ}$\xspace}

\hypersetup{pdfauthor={David van Treeck}, bookmarksnumbered=true, pdftitle={Influence of the source arrangement on  shell growth around GaN nanowires by molecular beam epitaxy}, colorlinks, citecolor=blue, linkcolor=blue, urlcolor=blue}

\begin{document}
\title{Influence of the source arrangement on  shell growth around GaN nanowires in molecular beam epitaxy}

\author{D. van Treeck} \email[Electronic mail: ]{treeck@pdi-berlin.de}
\affiliation{Paul-Drude-Institut für Festkörperelektronik, Leibniz-Institut im Forschungsverbund Berlin e.\,V., Hausvogteiplatz 5--7, 10117 Berlin, Germany}

\author{S. Fernández-Garrido}
\affiliation{Paul-Drude-Institut für Festkörperelektronik, Leibniz-Institut im Forschungsverbund Berlin e.\,V., Hausvogteiplatz 5--7, 10117 Berlin, Germany}
\affiliation{Grupo de Electrónica y Semiconductores, Dpto. Física Aplicada, Universidad Autónoma de Madrid, C/ Francisco Tomás y Valiente 7, 28049 Madrid, Spain.}

\author{L. Geelhaar}
\affiliation{Paul-Drude-Institut für Festkörperelektronik, Leibniz-Institut im Forschungsverbund Berlin e.\,V., Hausvogteiplatz 5--7, 10117 Berlin, Germany}

\begin{abstract}
In a combined experimental and theoretical study, we investigate the influence of the material source arrangement in a molecular beam epitaxy (MBE) system on the growth of nanowire (NW) core-shell structures. In particular, we study the shell growth of GaN around GaN template NWs under the boundary condition that Ga and N do not impinge on a given sidewall facet at the same time. Our experiments with different V/III ratios and substrate temperatures show that obtaining shells with homogeneous thickness along the whole NW length is not straightforward. Analyzing in detail the shell morphology with and without substrate rotation, we find that the different azimuthal angles of the sources have a major impact on the Ga adatom kinetics and the final shell morphology. In general, growth is possible only under directly impinging N, and Ga adatoms diffuse between the sidewalls and the top facet as well as the substrate, but not between adjacent sidewall facets. On the basis of these experimental results, we develop a diffusion model which takes into account different NW facets and the substrate. The model allows to describe well the experimental shell profiles and predicts that homogeneous shell growth can be achieved if the Ga and N source are arranged next to each other or for very high rotation speeds. Moreover, the modeling reveals that the growth on a given side facet can be categorized within one rotation in four different phases: the Ga wetting phase, the metal-rich growth phase, the N-rich growth phase, and the dissociation phase. The striking difference to growth processes on planar samples is that, in our case, diffusion takes place between different regions, i.e. the sidewall vs. the top facet and substrate, out of which on one N impinges not continuously, resulting in complex gradients in chemical potential that are modulated in time by substrate rotation. The comprehensiveness of our model provides a deep understanding of diffusion processes and the resulting adatom concentration, and could be applied to other 3D structures and material systems.
\end{abstract}

\maketitle

\section{INTRODUCTION}\label{sec:introduction}

One major advantage of core-shell nanowires (NWs) over planar structures is that the active region area can be drastically increased by simply increasing their aspect ratio. This makes core-shell structures particulary interesting for applications like LEDs or solar cells, where large active regions are beneficial \cite{Waag2011,LaPierre2013a}.
In order to grow well defined core-shell heterostructures, a precise control over the shell thickness along the whole length of the NW is needed. An epitaxial technique which allows the fabrication of heterostructures with monolayer precision is molecular beam epitaxy (MBE) \cite{Cho1999}. 
The molecular beams characteristic for this technique deposit material in direct line of sight from the source. Due to this directional nature, for three-dimensional (3D) structures like NWs, the chamber geometry plays\,--\,in contrast to conventional planar growth\,--\,a major role in the deposition process \cite{Foxon2009}.
Firstly, the position of the material sources in the chamber determines the polar angle $\alpha$ between the respective directions of the material beams and the substrate normal. As a consequence, the effective material fluxes on the NW top facet and the side facets may differ considerably.
Secondly, due to the circular arrangement of the material sources in the chamber, where each material source has a certain azimuthal angle $\beta$, the side facets are only exposed sequentially to the different material fluxes during substrate rotation.
Both geometrical aspects have a significant influence on the shell growth by MBE.

Among the enormous literature devoted to axial and radial growth of NWs by MBE, numerous studies mentioned the role of the source positions explicitly
\cite{Foxon2009,Glas_pssb_2010,Harmand_prb_2010,Galopin2011,Consonni2012,Dubrovskii2012b,Fernandez-Garrido2013a,Dubrovskii2013b,Hestroffer2013a,Bolshakov2014,Tersoff2015a,
Albert2015a,Gazibegovic2017,Kupers2018,Oehler2018,Lewis2018a}. 
It has been found that for the usual angles of $\alpha=30\text{--}40$\grad, gradients in the chemical potential lead to an increased adatom diffusion along the NW sidewalls towards the NW top facet or droplet.
For arsenide and nitride material systems, it has been shown that in addition to the NW dimensions and the V/III ratio, the polar angle $\alpha$ of the material sources has a significant influence on the final NW morphology \cite{Glas_pssb_2010,Galopin2011,Hestroffer2013a}.

The aspect that the side facets are only sequentially exposed to the different material beams has been widely ignored so far. 
Usually, only the fact that the flux impinging on the side facets can be lowered by self-shadowing has been considered by introducing a geometrical factor.
As a notable exception, Foxon et al.\,\cite{Foxon2009} pointed out for GaN NWs that the growth on the NW sidewalls may rather resemble migration enhanced epitaxy (MEE) than a classical MBE growth process. Hence, phenomena like Ga diffusion on the sidewall towards the NW top or substrate even before the side facet is exposed to the N beam might significantly affect the local growth rate and influence the thickness homogeneity of the shell.
Therefore, a more detailed understanding of the diffusion and growth processes on the NW side facets under changing conditions during rotation is needed for the controlled growth of homogeneous and complex core-shell structures.

In this study,
we investigate for GaN NWs, both experimentally and theoretically, the influence of the different azimuthal angles of the Ga and the N source on the Ga adatom kinetics on the NW and the resulting shell morphology.
We experimentally study the shell growth of GaN as a function of the V/III ratio and the temperature, where we analyze in detail the shell morphology with and without substrate rotation. On the basis of these experimental results, we develop a diffusion model which takes into account the different side and top facets of the NW as well as the substrate. The model allows to describe well the shell profile and conveys a deeper understanding of the adatom kinetics on the NW.
Investigating the adatom concentration and diffusion on the NW side facet during substrate rotation, we analyze in particular the influence of the circular arrangement of the material sources on the thickness homogeneity of the NW shell. Moreover, we discuss how the shell morphology can be controlled by optimizing the rotation speed and V/III ratio.

\section{SAMPLES, EXPERIMENTS, AND METHODS} \label{sec:methods}

We investigate five different core-shell NW ensembles A\,--\,E grown by plasma-assisted MBE. The GaN core NWs, also referred to as template NWs, were grown at a V/III ratio of 2.5 and a substrate temperature of about 780\celsius by means of self-assembly processes on a 3\,\mum thick Ti film sputtered on a Al$_2$O$_3$\,(0001) substrate. The Ti film was nitridized prior to the nucleation of the first NWs while ramping to the substrate temperature and a thin TiN layer formed. The template NWs have a mean diameter and length of about 35\,nm and 790\,nm, respectively, with a NW density of about 10$^9$\,cm$^{-2}$. Due to the rather low NW density almost all NWs are uncoalesced and mutual shadowing from the impinging fluxes is drastically reduced. More information on the template NW ensemble can be found elsewhere \cite{Wolz2015a,Treeck2018a,Calabrese2019}.
The above mentioned growth conditions for the template NW ensembles were used for all core-shell samples A\,--\,E.

The shell growth was performed directly after the growth of the template NWs and investigated as a function of the V/III ratio and the substrate temperature keeping the growth time constant at 35\,min. The different samples were grown at two different substrate temperatures, namely 600\celsius for samples A, B, C, and E and 640\celsius for sample D. The V/III ratio was changed from 0.98 to 1.4 by modifying the Ga flux $\Phi_\text{Ga}$ while keeping the N flux $\Phi_\text{N}$ constant at 9.2\,$\pm$\,0.5\,nm/min. 
The impinging fluxes $\Phi_\text{Ga}$ and $\Phi_\text{N}$ were calibrated in equivalent growth rate units of planar GaN layers as described elsewhere \cite{Heying2000}. 
The desorbing Ga flux during the experiments was monitored \textit{in situ} by line-of-sight quadrupole mass spectrometry (QMS) \cite{Koblmuller2004}.
The substrate temperatures were determined by a pyrometer which was calibrated for TiN as mentioned in Ref.\,\citenum{Treeck2018a}.
In our MBE system the Ga and the N cell have a polar angle $\alpha$ between the impinging beams and the substrate normal of 37.5\grad and are separated by an azimuthal angle $\beta$ of 144\grad. 
During the growth, samples A\,--\,D were continuously rotated in the direction in which the side facets are first exposed to the Ga beam and then, after a rotation of 144\grad, to the N beam. The rotation speed was 7 rounds per minute (rpm). Sample E was grown without substrate rotation.

In order to image the morphology of the core-shell NWs presented in this study, micrographs were recorded in a field-emission scanning electron microscope using an acceleration voltage of 5\,kV.  The length and diameter distributions, as well as the number density of the NW ensembles were determined by analyzing cross-sectional and top-view scanning electron (SE) micrographs with the help of the open-source software ImageJ \cite{Schneider2012a}.

The model to describe the shell growth on the NW side facets as well as the growth on the NW top facet and the substrate is based on a system of coupled one-dimensional (1D) diffusion equations which is solved by a numerical finite difference method (FDM) using the NumPy package \cite{numpy1,numpy2} for scientific computing with Python. The model is presented in detail in the paper and complemented in the supplemental information.

\section{RESULTS AND DISCUSSION}\label{sec:results}

\begin{figure*}[]
\includegraphics[width=0.95\columnwidth]{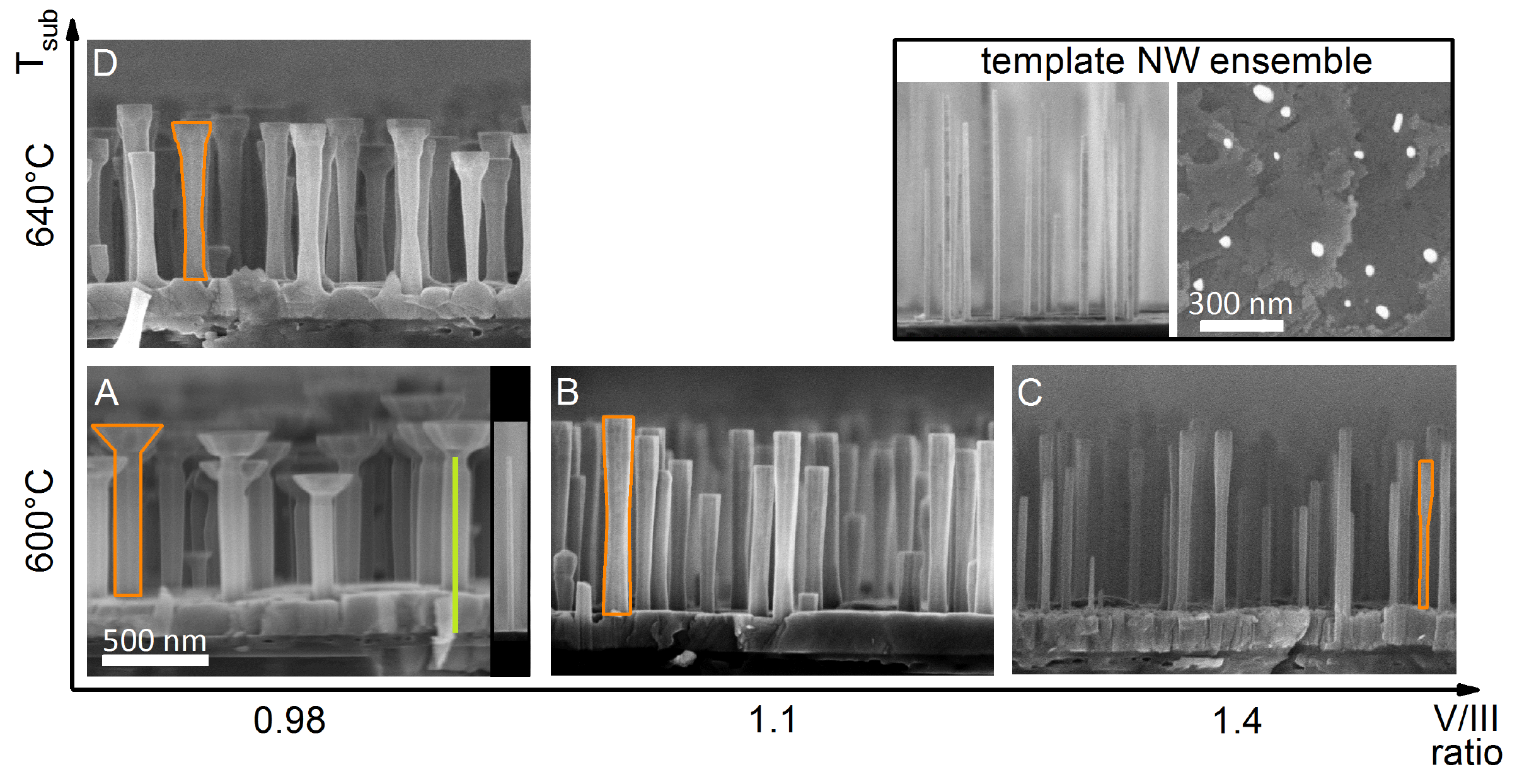}
\caption{SE micrographs showing the radial growth of GaN around GaN template NWs as a function of the V/III ratio (samples A, B, and C) and the substrate temperature (samples A and D). The orange outlines sketch the characteristic shapes of the respective NWs. The inset in the upper right corner shows a cross section and a plan view SE micrograph of the template NW ensemble, which was grown under conditions outside the range of this diagram. The inset in the SE micrograph of sample A depicts a template NW and the green rectangle next to the inset illustrates the extension of a template NW underneath a GaN shell.}
\label{fig1}
\end{figure*}

As a first step towards a comprehensive understanding of shell growth by MBE, in Fig.\,\ref{fig1} we investigate the radial growth of GaN around GaN template NWs as a function of the V/III ratio and the substrate temperature. 
The inset in the upper right corner of the growth map shows a cross section and a plan view SE micrograph of the template NW ensemble.
For purposes of direct comparison, the inset on the right-hand side of Sample A shows an SE micrograph of a single average template NW, and the green rectangle next to the inset illustrates the extension of a template NW underneath a GaN shell. The SE micrograph of sample A depicts that a substrate temperature of 600\celsius and nominally metal-rich growth conditions with a V/III ratio of 0.98 result in a homogeneous shell along the whole NW length, however, also lead to the formation of wide, trapezoid-shaped top segments. It should be noted that at 600\celsius there was no Ga adatom desorption detected in the QMS. 
Increasing the V/III ratio to 1.1, hence, slightly N-rich growth conditions, NWs with straight top segments are obtained as shown for sample\;B. However, the radial growth along the NW side facets is not homogeneous anymore. Increasing radial growth towards the NW top and the NW bottom leads to a pronounced hourglass shape of the NWs. Also for much higher V/III ratios as for sample C this characteristic shape of the shell is maintained, whereas due to the reduced Ga flux less material is grown. 

Increasing the substrate temperature for sample D with regard to sample A leads to desorption of Ga adatoms (detected by QMS) which in turn results in N-rich conditions and a similar hourglass shape of the shell. The more pronounced top segments in comparison to sample B are likely the result of an increased adatom diffusion towards the NW top at higher temperatures \cite{Songmuang2010,Fernandez-Garrido2013a}. 
Furthermore, we note that growth does not only take place on the side and top facets of the template NWs. All samples exhibit a parasitic GaN layer that has grown on the TiN substrate at the foot of the NWs as a result of the low substrate temperature. Such a parasitic GaN layer does not form at elevated substrate temperatures as in the case of the template NW ensemble (see inset) \cite{Wolz2015a,Treeck2018a,Calabrese2019}.

In general, it was not possible to obtain a homogeneous shell along the whole length of the template NWs at low temperatures while avoiding the formation of wide top segments.
A widening at the NW top is well known to occur for metal-rich growth conditions and has been attributed to preferential nucleation at the edge between side and top facet due to the high adatom concentration, with the formation of stacking faults and cubic insertions possibly being another relevant factor for lateral growth \cite{Songmuang2010,Fernandez-Garrido2013a,Zhang2016u,Gruart2019}.
The homogeneous shell obtained for Ga-rich growth conditions is most likely the result of a continuous wetting of the side facets during the growth; i.\,e.~the Ga on the side facets is never fully consumed when the facets are exposed to the N beam and growth takes place. However, despite the homogeneous shell, the NW morphology obtained for metal-rich growth with its wide top segments is not suitable for a further device processing of the NW ensemble. Moreover, the side facets are shadowed more and more from the impinging beams by the top segments which renders the growth of more complex shell structures impossible.
Hence, the best result in terms of shell homogeneity was obtained for sample B grown at 600\celsius under nominally N-rich growth conditions, where no trapezoid-shaped top segments are formed.\\

\begin{figure}[t!]
\includegraphics[width=0.92\columnwidth, trim= 0 0 0 0]{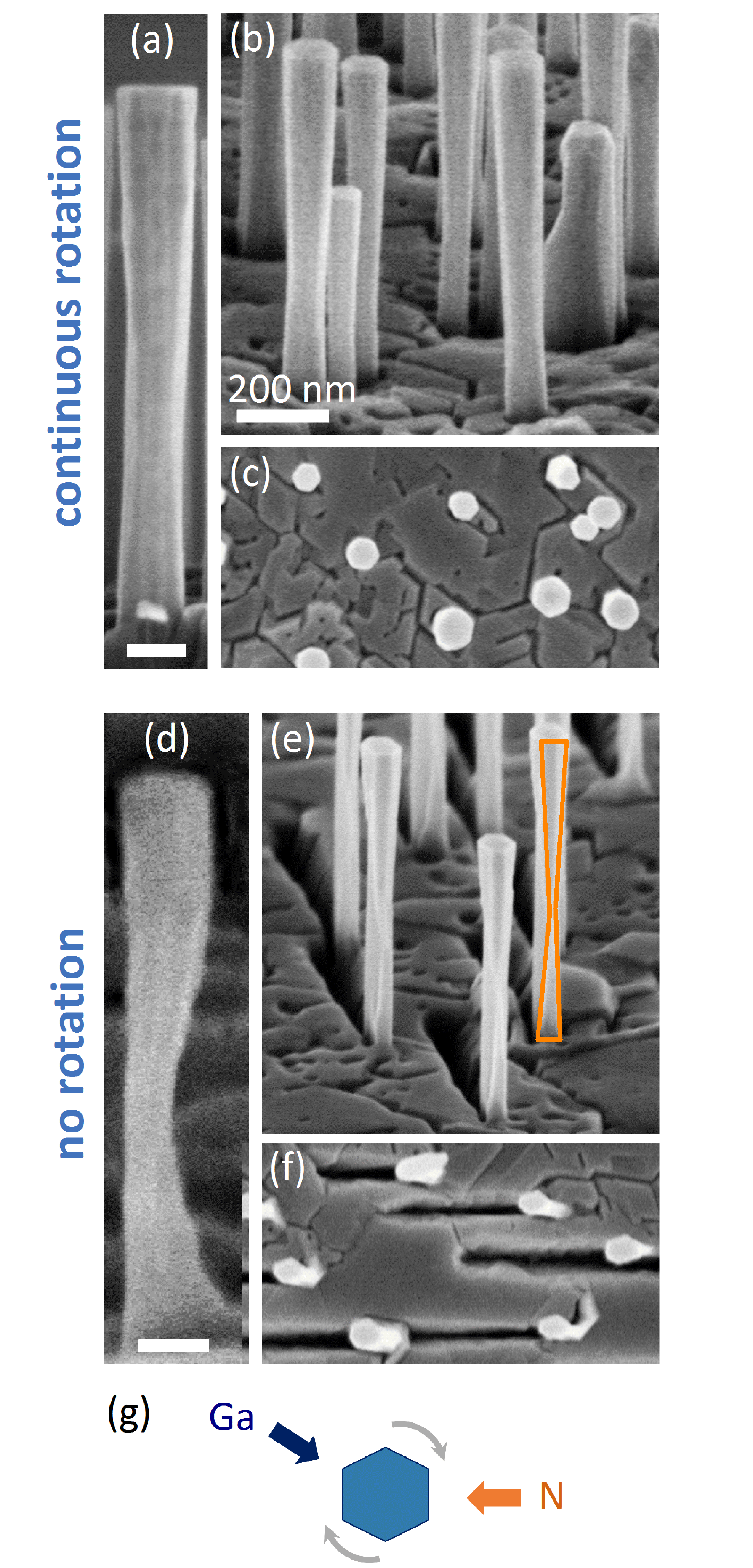}
\caption{The SE micrographs show (a)\,--\,(c) sample B  grown \textit{with} continuous substrate rotation and (d)\,--\,(f) sample E grown \textit{without} substrate rotation. The cross section micrographs (a) and (d) depict typical single NWs from these ensembles where the scale bars correspond to 100\,nm. The bird's eye (b/e) and plan view (c/f) micrographs give a representative overview of both samples. The orange outline in (e) sketches the double half cone shape of the shell of sample E. Note that the perspective of micrograph (e) is rotated with respect to (d) and (f) to give a better impression of the shell morphology in the radial direction. The direction of the impinging N is revealed by the grooves in the parasitic GaN layer seen in (e) and (f). The micrographs (b), (c), (e) and (f) are to scale and (b) and (e) were measured at a tilt angle of 20\grad and 25\grad with respect to the substrate, respectively. The plan view sketch (g) shows the source geometry of our MBE system, where the Ga and the N beam are separated by an azimuthal angle $\beta$ of 144\grad. The grey arrows depict the rotation direction of sample B.} 
\label{fig2}
\end{figure}

In order to better understand the mechanisms leading to the formation of an hourglass shape of the GaN shell for various growth conditions, in Figs.\,\ref{fig2}(a)\,--\,(f) we compare sample B grown with continuous substrate rotation to sample E grown without substrate rotation at otherwise similar growth conditions. 
The goal of analyzing core-shell structures grown without substrate rotation is to learn more about the adatom diffusion processes on the different NW facets. Since in our MBE system, the Ga and the N beam are, with an azimuthal angle $\beta$ of 144\grad, opposed to each other as shown in the sketch in Fig.\,\ref{fig2}\,(g), shell growth without substrate rotation can only take place if there is diffusion of adatoms.

The NW depicted in Fig.\,\ref{fig2}\,(a) shows the characteristic tapering found for the optimum sample B with bottom, center, and top diameters of 121\,nm, 109\,nm, and 146\,nm, respectively, and a length of 920\,nm. 
Figs.\,\ref{fig2}\,(b) and (c) are representative SE micrographes of the NW ensemble. The NWs are almost completely uncoalesced and have a hexagonal faceting as it can be seen in the plan view image (c). For some NWs, mutual shadowing of adjacent NWs cannot be excluded, however, we found that also isolated NWs which where not shadowed by any other NWs in their surrounding showed a similar hourglass morphology. Hence, mutual shadowing of the NWs cannot be the decisive mechanism responsible for this characteristic shell shape.   

Analyzing the NW morphology of sample E in Fig.\,\ref{fig2}\,(d), one can directly see that the growth of an hourglass-shaped shell also appears for the non-rotated sample, however, only on the N exposed side (right hand side in the image). The NW length is about 780\,nm and the diameters along the NW from bottom to center to top are 145\,nm, 62\,nm, and 119\,nm, respectively. Hence, with the template NWs having a mean diameter of about 35\,nm, there is only very little radial growth in the NW center. This phenomenon can also be seen in Fig.\,\ref{fig2}\,(e), where the shell seems to have a kind of double half cone shape with less radially grown material towards the NW center. 
Furthermore, the shadowing of the N beam by the NWs leads to deep grooves in the parasitic GaN layer clearly visible in the plan view micrograph shown in Fig.\,\ref{fig2}\,(e). It should be noted that most NWs which have an hourglass-shaped shell are not affected by shadowing by a neighboring NW, which otherwise would be indicated by a groove crossing the NW foot. 

In general, the fact that shell growth only takes place at the N exposed side of the NW implies that N diffusion on the NW facets is negligible, in agreement with literature \cite{Koleske1998,Lymperakis2009,Fernandez-Garrido2013a}. 
The absence of N adatom diffusion also explains the groove formation in the parasitic GaN layer.
Moreover, that radial growth is much less pronounced in the NW center, indicates that Ga adatom diffusion around the NWs from m-plane to m-plane is strongly suppressed for these growth conditions. Hence, the Ga adatoms contributing to the growth on the N exposed side have to diffuse there either from the top facet or the substrate, which eventually results in the observed hourglass shape of the shell.  
The findings for sample E suggest that also for the rotated sample B, where the NWs show a less pronounced but similar hourglass shape, the inhomogeneity of the shell is mainly caused by Ga adatom diffusion processes and at most to a small extent by directly impinging Ga. 
In other words, the fact that the anion and cation beams impinge on different NW sides plays a major role for the final shell morphology.

\begin{figure}[t!]
\includegraphics[width=0.85\columnwidth]{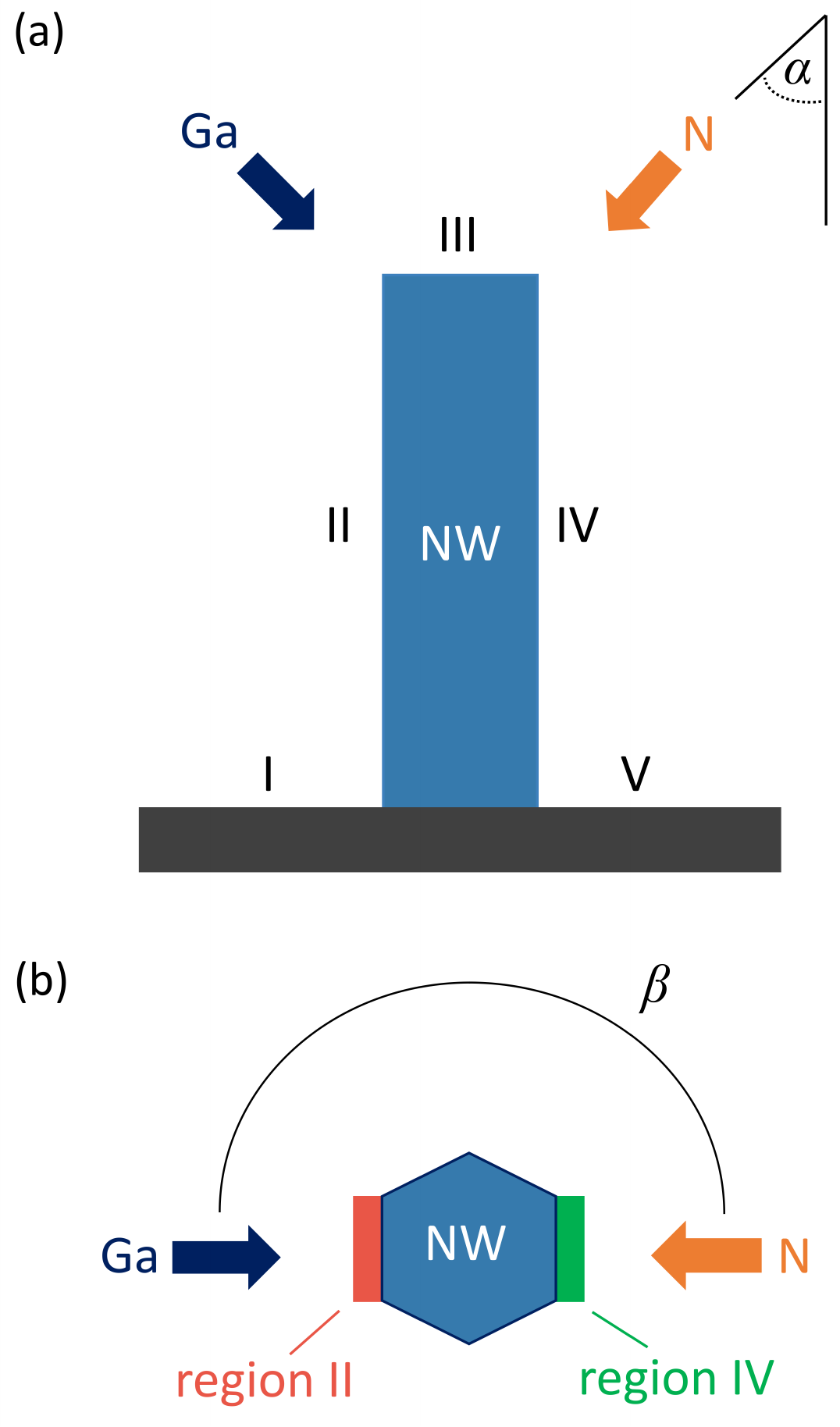}
\caption{(a) The cross section sketch of a NW depicts the five different regions distinguished in our model: (I) the substrate, (II) the NW facet which is directly exposed to the Ga flux, (III) the NW top facet, (IV) the NW facet which is directly exposed to the N flux, and (V) again the substrate. The polar angle $\alpha$ is the angle between the flux direction of the material sources and the substrate normal. (b) The top view sketch of the NW shows the azimuthal arrangement of the side facet regions II and IV as well as of the Ga and the N source which are separated by an azimuthal angle $\beta$.}
\label{fig3}
\end{figure}

In order to elucidate in more depth how the arrangement of the material sources in an MBE system influences adatom diffusion and thus the final NW shape, we developed a 1D diffusion model for the Ga adatom concentration on the NW. The model takes into account five different regions as sketched in Fig.\,\ref{fig3}\,(a). 
In general, we assume that the diffusion of N is negligible as indicated by the analysis of sample E in Fig.\,\ref{fig2}. Hence, in the model we only consider the diffusion of Ga adatoms. Moreover, we assume that in the presence of Ga on a given facet, each N adatom impinging on the facet contributes to the growth. For metal-rich growth conditions, the growth rate is thus limited by the impinging N flux. It should be noted that in agreement with our experimental findings above, inter-diffusion of Ga adatoms between the different side facets (region II and IV in the model) is neglected. 
In the following we will introduce the model for the two different scenarios, \textit{with} and \textit{without} substrate rotation, and apply it to the shape occurring for samples B and E, respectively. 
Since experimentally we did not observe any significant Ga desorption for the samples B and E, Ga adatom desorption is not considered in the model.

\subsection{Modeling shell growth without substrate rotation}
\label{sec:A}

In the case \textit{without} substrate rotation, the regions in Fig.\,\ref{fig3}\,(a) can be attributed to (I) the substrate, (II) the NW facet which is directly exposed to the Ga flux, (III) the NW top facet, (IV) the NW facet which is directly exposed to the N flux, and (V) again the substrate. The angle $\alpha$ is the angle between the impinging beams and the substrate normal and is the same for both sources. The plain-view sketch of Fig.\,\ref{fig3}\,(b) illustrates the two different side facet regions II and IV. The Ga and N source and hence region II and IV are separated by an azimuthal angle $\beta$ which depends on the chamber geometry of the MBE system. 
The Ga adatom concentrations $n$ of the different regions (I)\,--\,(V) are described by the following 1D diffusion equations:

\renewcommand{\labelenumi}{(\Roman{enumi})}
\begin{enumerate}
\item $\begin{aligned}[t]
    \frac{\partial n_\text{I}(x_\text{I},t)}{\partial t} &=- D_\text{sub} \frac{\partial^2 n_\text{I}(x_\text{I},t)}{\partial x_\text{I}^2} + \frac{n_\text{I}(x_\text{I},t)}{\tau_\text{sub}} + \chi_\parallel J_\text{Ga}
\end{aligned}$
\item $\begin{aligned}[t]
    \frac{\partial n_\text{II}(x_\text{II},t)}{\partial t} &= D_\text{side} \frac{\partial^2 n_\text{II}(x_\text{II},t)}{\partial x_\text{II}^2} +  \chi_\perp J_\text{Ga} 
\end{aligned}$
\item $\begin{aligned}[t]
    \frac{\partial n_\text{III}(x_\text{III},t)}{\partial t} &= D_\text{top} \frac{\partial^2 n_\text{III}(x_\text{III},t)}{\partial x_\text{III}^2} + \frac{n_\text{III}(x_\text{III},t)}{\tau_\text{top}} + \chi_\parallel J_\text{Ga} 
\end{aligned}$
\item $\begin{aligned}[t]
    \frac{\partial n_\text{IV}(x_\text{IV},t)}{\partial t} &= D_\text{side} \frac{\partial^2 n_\text{IV}(x_\text{IV},t)}{\partial x_\text{IV}^2} + \frac{n_\text{IV}(x_\text{IV},t)}{\tau_\text{side}} 
\end{aligned}$
\item $\begin{aligned}[t]
    \frac{\partial n_\text{V}(x_\text{V},t)}{\partial t} &= D_\text{sub} \frac{\partial^2 n_\text{V}(x_\text{V},t)}{\partial x_\text{V}^2} + \frac{n_\text{V}(x_\text{V},t)}{\tau_\text{sub}} + \chi_\parallel J_\text{Ga} 
\end{aligned}$
\end{enumerate}

\noindent
where $D_\text{sub}$, $D_\text{side}$, and $D_\text{top}$ are the diffusion coefficients and $\tau_\text{sub}$, $\tau_\text{side}$, and $\tau_\text{top}$ are the lifetimes of Ga adatoms on the substrate, the side facet, and the top facet of the NW, respectively. In our model these six parameters describing the diffusion terms $D~{\partial^2 n}/{\partial x^2}$ and the incorporation terms ${n}/{\tau}$ are considered as fitting parameters. We would like to point out that, in general, the diffusion coefficients $D$ may be affected by the Ga/N ratio. In our model, we assume that all diffusion coefficients are constant within one region and in time. Hence, the parameter values of the diffusion coefficients are mean values in time and space for the respective region.

The different contributions of the impinging Ga flux for facets oriented parallel and perpendicular to the substrate are described by the flux terms $\chi_\parallel J_\text{Ga}$ and $\chi_\perp J_\text{Ga}$, respectively, where $\chi_\parallel = \cos \alpha$ and $\chi_\perp = \sin \alpha$.
Since in this section, we assume that the Ga and N species do not impinge on a given sidewall facet at the same time, the equations (I)\,--\,(V) are only valid for azimuthal angles in the range 120\grad\,$<\beta<$\,240\grad. A more general model will be introduced in section\;\ref{sec:B}.
Moreover, it should be noted that for region I, we did not consider the shadowing from the N beam by the NW, since the characteristics of this region are mainly governed by the non-shadowed area. We will discuss this in more detail below.
The spatial ranges for the different regions are defined as 
\begin{equation}
\label{eq:growth rate}
\begin{aligned}
x_\text{I,V} &\in \text{[0,} ~\infty\text{]}\\
x_\text{II,IV} &\in \text{[}0, l_\text{eff}(t)\text{]}\\
x_\text{III} &\in \text{[}0, d(t)\text{]}  
\end{aligned}
\end{equation}
with the effective NW length $l_\text{eff}(t)=l(t)-h_\text{para}(t)$ between parasitic layer and NW top facet, where $l(t)$ is the total length of the NW starting from the substrate and $h_\text{para}(t)$ is the height of the parasitic layer. The parameter $d(t)$ is the top facet diameter starting at the diameter of the template NW.
Regarding the boundary conditions, the adatom concentration $n$ as well as the diffusion fluxes are chosen to be continuous at the boundaries between the different regions j (with j $\in$ I, ..., V ) for all times $t$:

\begin{equation}
\label{eq:BC}
\begin{aligned}
&n_\text{I}(\infty,t)=n_\infty,\\
&n_\text{I}(0,t)=n_\text{II}(0,t)
,~~~~~~~~~~~~~~~~~~\frac{\partial n_\text{I}}{\partial x_\text{I}}|_{x_\text{I}=0} = - \frac{\partial n_\text{II}}{\partial x_\text{II}}|_{x_\text{II}=0},\\
&n_\text{II}(l_\text{eff}(t),t)=n_\text{III}(0,t)
,~~~~~~~~~~\frac{\partial n_\text{II}}{\partial x_\text{II}}|_{x_\text{II}=l_\text{eff}(t)} = \frac{\partial n_\text{III}}{\partial x_\text{III}}|_{x_\text{III}=0},\\
&n_\text{III}(d(t),t)=n_\text{IV}(l_\text{eff}(t),t)
,~~~~~\frac{\partial n_\text{III}}{\partial x_\text{III}}|_{x_\text{III}=d(t)} = -\frac{\partial n_\text{IV}}{\partial x_\text{IV}}|_{x_\text{IV}=l_\text{eff}(t)},\\
&n_\text{IV}(0,t)=n_\text{V}(0,t)
,~~~~~~~~~~~~~~~~\frac{\partial n_\text{IV}}{\partial x_\text{IV}}|_{x_\text{IV}=0} = -\frac{\partial n_\text{V}}{\partial x_\text{V}}|_{x_\text{V}=0},\\
&n_\text{V}(\infty,t)=n_\infty,
\end{aligned}
\end{equation}

\noindent hence, we assume equality of the chemical potentials at the boundaries and the conservation of the number of adatoms in the whole system. For region I and V, $n$ goes to an equilibrium value of $n_\infty$ far away from the NW. 
For the initial conditions, the adatom concentration $n_\text{j}$ is chosen to be zero for all $x_\text{j}$ in all regions j:
\begin{equation}
\label{eq:IC}
n_j(x_\text{j},0)=0 
\end{equation}

Since the Ga adatom concentration on the NW during the growth is experimentally not accessible, one has to model the final shell morphology, i.\,e.~the grown material, in order to learn more about the adatom concentration and kinetics.
The thickness of the grown material $GM_\text{j}$ at a certain point $x_\text{j}$ of the different regions $j$ is described by integrating the growth rate $GR_\text{j}={n_\text{j}(x_\text{j},t)}/{\tau_\text{j}}$ (incorporated material per time) over the total growth time $t_\text{g}$:
\begin{equation}
     GM_\text{j}(x_\text{j})= \int_{0}^{t_{g}} \frac{n_\text{j}(x_\text{j},t)}{\tau_\text{j}}~dt   ~~~~~\text{for~~j} \in \text{I, .., V}
\label{eq_GM}
\end{equation}

According to our assumptions, the growth rate $GR_\text{j}$ in each region can only be as high as the impinging N flux, which can be expressed as ${n_\text{j}(x_\text{j},t)}/{\tau_\text{j}}\le\chi_\text{j} J_N$.
Hence, only one Ga adatom per impinging N atom is incorporated, excess adatoms are free to further diffuse.\\

The modeling of the grown material for the shell, the top segment and the parasitic layer on the substrate can be achieved according to Eq.\,(\ref{eq_GM}) by adjusting the fitting parameters $D$ and $\tau$ in the different regions. 
It should be noted that in order to get a good description of the shell growth, we took into account that the growth of the top facet and the parasitic layer affects the section of the side facet where growth can take place; i.\,e.~once the parasitic layer increases, the lower part of the NW side wall is covered and once the top segment grows longer the side facet is enlarged.

The shell profile which develops on the side facet is mainly described by the adatom diffusion length on the side facet $\lambda_\text{side}=\sqrt{D_\text{side}\tau_\text{side}}$, where $D_\text{side}$ mainly determines the symmetry of the shell along the NW and $\tau_\text{side}$ regulates the shell thickness. 
The boundary areas at the very top and bottom part of the side facet, are also strongly influenced by the parameters $D_\text{top}$/$D_\text{sub}$ and $\tau_\text{top}$/$\tau_\text{sub}$, which mainly characterize the Ga adatom diffusion onto the side facet.
Playing with these parameters allows to lower or increase the adatom concentration at the boundaries to region IV and thus the thickness of the shell at the bottom and the top of the side facet. 
However, the variation of $D$ and $\tau$ in the regions III and IV is restricted, since also the modeling of the experimentally observed thickness of the top segment and the parasitic layer is taken into account and is essential for a reasonable description of the shell growth. 
As a reference for the modeling of the thickness of the parasitic layer, we used 227.5\,nm which corresponds to the value obtained for Ga limited growth with $\chi_\parallel\Phi_\text{Ga}=6.5~\text{nm/min}$ and $t_\text{g}=35~\text{min}$ far away from NWs, i.\,e.~excluding shadowing effects of adjacent NWs. 
Since a decrease of the layer thickness towards the NW was usually not observed in SE micrographs, we assumed the thickness to be close to 227.5\,nm for distances larger than 100\,nm from the NW. 
Due to the smooth shape of the NWs, it was not possible to identify the onset for the top segment growth from SE micrographs which rendered an estimation of the top segment thickness impossible. Therefore, for the modeling we assumed the mean height of the top segments to be between 227.5 (value explained above) and 255.5\,nm, where the latter thickness is obtained for N-limited growth at the top facet, hence, the maximum possible thickness. 

Due to the fact that all regions are interconnected and depend on each other in a particular way, the whole system, i.\,e.~a certain configuration of shell, top segment, and parasitic layer thickness, can only be described by a unique set of the fitting parameters. Hence, a good modeling of the experimentally observed thicknesses can be achieved only within a small parameter window. 
Regarding the sensitivity of the fits with respect to a change of the different fitting parameters, we found that varying the parameters $\tau_\text{top}$, $\tau_\text{side}$, $\tau_\text{sub}$ and $D_\text{side}$ by about ±10\% around the optimum value, results in a deviation of the modeled shell profile of about ±5\%. The dependence of the fits on the parameters $D_\text{top}$ and $D_\text{sub}$ is comparatively less sensitive. For these parameters, a variation of about ±85\% results in a deviation of ±5\% of the modeled shell profile.
More details on the behavior and sensitivity of the fits as well as on the general fitting procedure can be found in the supplemental information.
We would like to emphasize that with this model, our main object is to develop a qualitative understanding of the complex diffusion processes on the NW rather than extracting precise values for the parameters $D$ and $\tau$ in the different regions. 
Once a good description of the shell thickness, the top segment and the parasitic layer is achieved, the model allows to analyze the adatom concentration in the different regions for all times $t \in [0, t_\text{g}$].\\

\begin{table}[]
 \caption{Fitting parameters for sample E}
 \label{tab1}
\begin{ruledtabular}
  \begin{tabular}{c c c c c}
sample 	& region  		& $D$ (cm$^2$/s) 		& $\tau$ (s) 	& $\lambda$ (nm) 	\\
\hline
E		& top (III) 	& 2.5$\times$10$^{-12}$ 	& 0.25 			& 8   				\\
		& side (II+IV) 	& 4.5$\times$10$^{-10}$ 	& 0.45 			& 142  				\\
		& sub (I+V) 	& 3.5$\times$10$^{-12}$ 	& 0.6 			&  14				\\	
\end{tabular}
\end{ruledtabular}
\end{table}

\begin{figure}[]
\includegraphics[width=0.99\columnwidth]{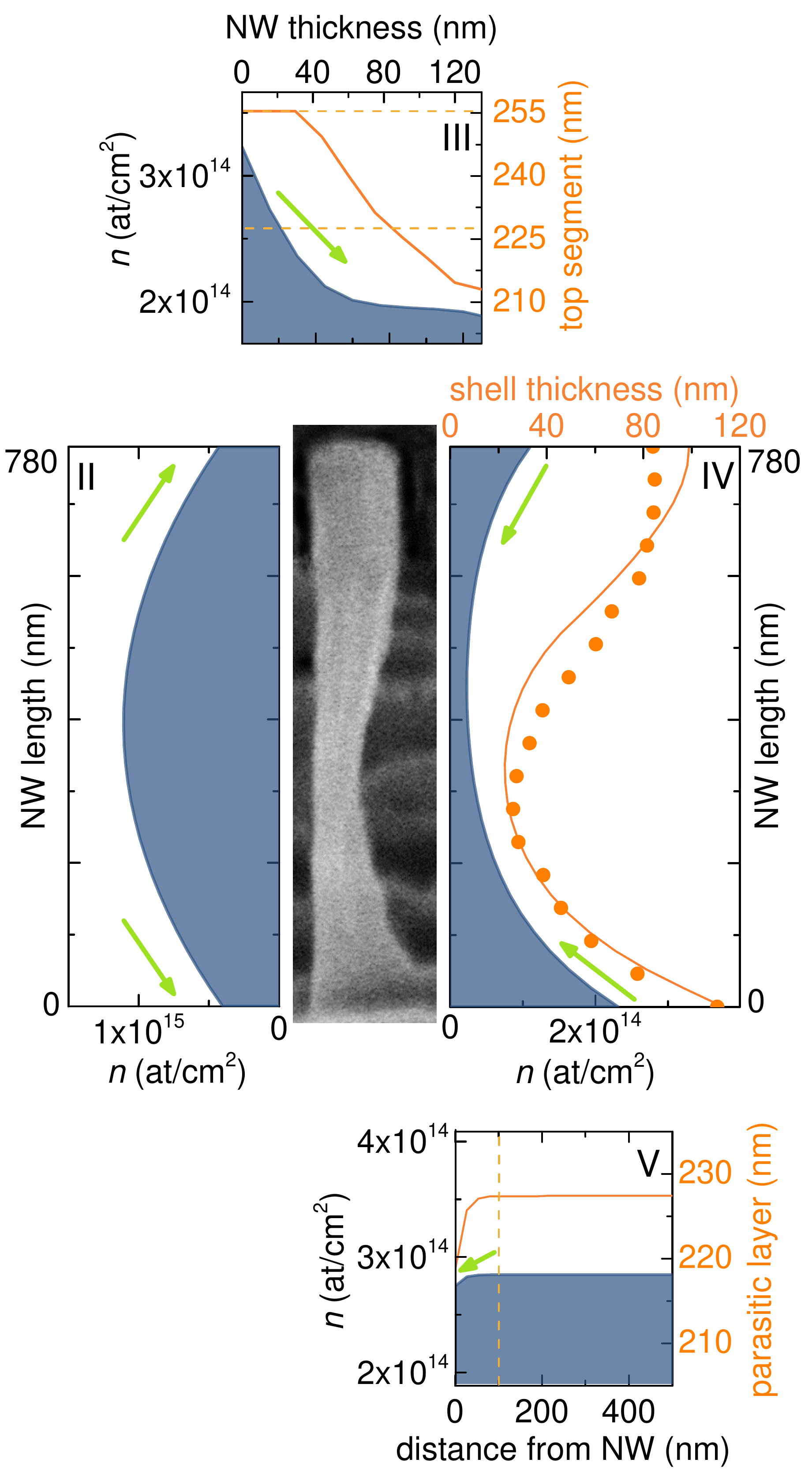}
\caption{The different graphs show the modeled Ga adatom concentration $n$ (blue) and the thickness of the grown material $GM$ (orange) according to Eq.\,(\ref{eq_GM}) on the side facet (region IV), the top facet (region III), and the substrate (region V) for a representative NW of sample E for $t=t_\text{g}$. The micrograph in the center depicts the modeled NW. The dashed lines in the graph of region III at 255.5 and 227.5\,nm mark the thickness obtained for N-limited and Ga-limited growth, respectively. The dotted data in the graph of region IV is the experimentally determined shell thickness of sample E. The dashed line in the graph of region V at 100\,nm marks the distance to the NW above which we assume a constant thickness of the parasitic layer. The green arrows indicate the mean diffusion direction of Ga adatoms.}
\label{fig4}
\end{figure}

Fig.\,\ref{fig4} shows the modeled grown material as well as the adatom concentrations at $t=t_\text{g}$ of the different regions II, III, IV and V of a representative NW of sample E. 
The respective fitting parameters are given in Table\,\ref{tab1}. 
The angle $\alpha$ was chosen according to our MBE system (see experimental section) and the angle $\beta=$\,144\grad is within the range in which the equations (I)\,--\,(V) are valid.
Analyzing the adatom concentration of region II, one finds that up to about one monolayer (ML) of Ga accumulates on the Ga exposed side facet of region II (m-plane: 1 ML = 1.21$\times$10$^{15}$\,at/cm$^2$) and that the adatom concentration decreases towards the substrate (region I) and towards the NW top facet (region III). Since region II is only exposed to Ga there is no growth on this side facet.
The adatom concentration on the top facet has a value of around 0.3 ML close to region II (c-plane: 1 ML = 1.14$\times$10$^{15}$\,at/cm$^2$) and decreases to about 0.1 ML at the boundary to region IV. A similar behavior is observed for the thickness of the top segment from one to the other boundary. Close to the Ga exposed side facet (region II), the segment reaches the maximum thickness of 255.5\,nm. In contrast, close to the N exposed side facet (region IV), the thickness is 213\,nm. 
It should be noted that a difference in thickness of the top segment from one to the other side was not observed experimentally by analyzing SE micrographs of single NWs of sample E. The average thickness of the top segment is 236\,nm.
Regarding the N exposed side facet (region IV), the adatom concentration has its maximum values at the boundaries to the top region III and the substrate region V with values of about 0.1 ML and around 0.2 ML, respectively.  In the NW center, the adatom concentration is much lower than at the edges. The modeling of the shell thickness of region IV shows that the experimentally observed tapered shell profile (dots) along the NW of the non-rotated sample E can be well described by our model (line). 
The Ga adatom concentration on the substrate far away from the NW is with about 0.25 ML between the two extreme values of the top facet. The grown layer thickness is around 227.5\,nm which coincides well with the experimentally found layer thickness in non-shadowed areas and only drops visibly for distances smaller than about 50\,nm.

In general, the belly shape of the adatom concentration in region II can be explained by the diffusion of Ga adatoms towards the substrate and the top of the NW, as indicated by the green arrows. The driving force for these diffusion processes are the different chemical potentials in the regions I, II, and III resulting from different material incorporation rates (growth rates) and impinging fluxes.
For instance, without any additional diffusion fluxes one would expect region III to have a similar adatom concentration as the substrate regions I and V far away from the NW, since for both regions growth takes place in c-plane direction and both are exposed to the same Ga and N fluxes. However, as a consequence of the continuous Ga supply in region II and the nominally N-rich growth conditions in region III resulting in a gradient in the chemical potential, Ga adatoms diffuse onto the top facet, which even leads to a Ga excess and hence N-limited growth near the boundary of the two regions.
In contrast, the adatom concentration is decreased towards region IV which results from the low chemical potential on the N exposed side facet IV and leads to a segment thickness below 227.5\,nm, the one expected for growth at the rate of the impinging Ga flux. 
Hence, on this side of the NW, the gradient in the chemical potential causes Ga adatoms to diffuse from the top facet and the substrate onto the side facet of region IV and towards the NW center. The resulting profile of the adatom concentration resembles the characteristic hourglass shape which according to Eq.\,(\ref{eq_GM}) manifests itself in the final shell profile. It should be mentioned that the flattening shell profile towards the NW top results from the shift of $l_\text{eff}(t)$ along the NW with increasing total NW length $l(t)$ and parasitic layer height $h(t)$. 
Moreover, the top facet diameter $d(t)$ changes from 35\,nm (diameter of template NW) to about 120\,nm (template NW + shell) leading to a decrease in the diffusion towards region IV with increasing growth time.

Regarding the fitting parameters shown in Table\,\ref{tab1}, we already mentioned that the intention of our model is not to extract precise values for $D$ and $\tau$ but to qualitatively understand the adatom kinetics for the shell growth. Nevertheless, we would like to address a few points.
In a realistic scenario, it can be assumed that there are diffusion barriers at the boundaries between the different regions \cite{Dubrovskii2014a}, e.\,g.~because the surfaces of the regions have different crystallographic orientations.
Since in our model we did not explicitly consider any diffusion barriers at the boundaries of the regions, their impact on the diffusion from region to region might be included in our fitting parameters. 
In general, diffusion barriers on the surface affect the diffusion coefficient which is defined by $D=D_\text{0}~\exp(-E_\text{diff}/k_\text{B} T)$ where $D_\text{0}$ is the temperature-independent pre-factor, $k_\text{B}$ is the Boltzmann factor, and $E_\text{diff}$ is the height of the potential barriers on the surface. 
Analyzing the fitting parameters of table\,\ref{tab1}, one finds that the values of $D_\text{top}$ and $D_\text{sub}$ of the c-plane regions which result in a good description of the shell profile are about two orders of magnitude lower than $D_\text{side}$ of the m-plane region IV. 
Considering the fact that the Ga adatoms which contribute to the growth of the shell have to diffuse from the top facet and the substrate onto the side facet, processes which are determined by $D_\text{top}$ and $D_\text{sub}$, it is indeed likely that for these diffusion processes, the barrier effects at the boundaries are included in $D_\text{top}$ and $D_\text{sub}$ and not in the parameters $D_\text{side}$ and $\tau_\text{side}$ describing the diffusion on the side facet.
Hence, the diffusion coefficients $D_\text{top}$ and $D_\text{sub}$ are likely dominated by the barrier heights at the boundaries and have to be seen as effective diffusion coefficients rather than realistic values for the respective regions. 
It should be noted that a difference of a few hundred meV in barrier height would be enough to explain the observed two orders of magnitude difference in $D$ between the side facet and the top facet/substrate.
According to density-functional theory calculations of Lymperakis \textit{et al.}\,\cite{Lymperakis2009}, it is rather unlikely that the differences between $D_\text{side}$ and $D_\text{top/sub}$ result from the different barrier heights directly on the m- or c-plane since the barrier heights on the m-plane are actually higher, which would result in a comparatively smaller diffusion coefficient $D_\text{side}$.
Moreover, if the barrier heights at the boundaries dominate the values of $D_\text{top}$, we might underestimate the diffusion length of the Ga adatoms on the top facet in our model. A larger diffusion length would result in a less pronounced thickness difference of the top segment from one to the other side, which would explain why the SE micrographs of sample E do not show recognizable change in thickness of the top segment.
The only values for $D$ found in literature are with 2$\times$10$^{15}$\,cm$^2$/s at 740\celsius on the m-planes\,\cite{Brandt2004} and about 1.1$\times$10$^{15}$\,cm$^2$/s at 700\celsius on the c-planes\,\cite{Koleske1998} several orders of magnitude lower than the values we obtained, but due to very different experimental conditions not directly comparable to our results. 

Regarding the Ga adatom lifetimes $\tau$, the values for the different regions are all in the hundred ms range with variations of about a factor two. Similar values have been found analyzing planar GaN growth by metal organic vapor deposition using a kinetic growth model \cite{Koleske1998}.
It should be mentioned that modeling the shell growth for much larger Ga adatom lifetimes (two orders of magnitude or more) leads to an accumulation of several tens of ML of Ga in the different regions, increasing with further growth time. Since such a scenario is not realistic, the values we obtained for $\tau$ seem to be reasonable.

Since $D$ and $\tau$ are hard to determine experimentally, it is generally more revealing to compare the corresponding diffusion length $\lambda$ obtained for the respective regions. For Ga adatom diffusion on the m-polar side facets of NWs, diffusion lengths of about 40 to 100\,nm were reported to be reasonable for high temperature growth (around 800\celsius) \cite{Debnath_apl_2007,Lymperakis2009,Galopin2011,Consonni2012}. The higher Ga adatom diffusion length of 144\,nm we found for sample E can be explained by the absence of Ga desorption, which is the limiting factor at high temperatures \cite{Galopin2011}.

From modeling sample E, we can conclude that in the extreme case without rotation, indeed, the different azimuthal angles of Ga and N causing adatom diffusion from the top and the substrate onto the N exposed region seem to be mainly responsible for the inhomogeneous shape of the shell.
However, the question is now, whether these conclusions also hold for the scenario \textit{with} substrate rotation which is necessary to obtain a radially homogeneous shell and, moreover, whether a homogeneous shell thickness along the NW can be achieved for a different arrangement of the material sources or different growth conditions.

\subsection{Modeling shell growth with substrate rotation}
\label{sec:B}

\begin{figure}[]
\includegraphics[width=0.96\columnwidth]{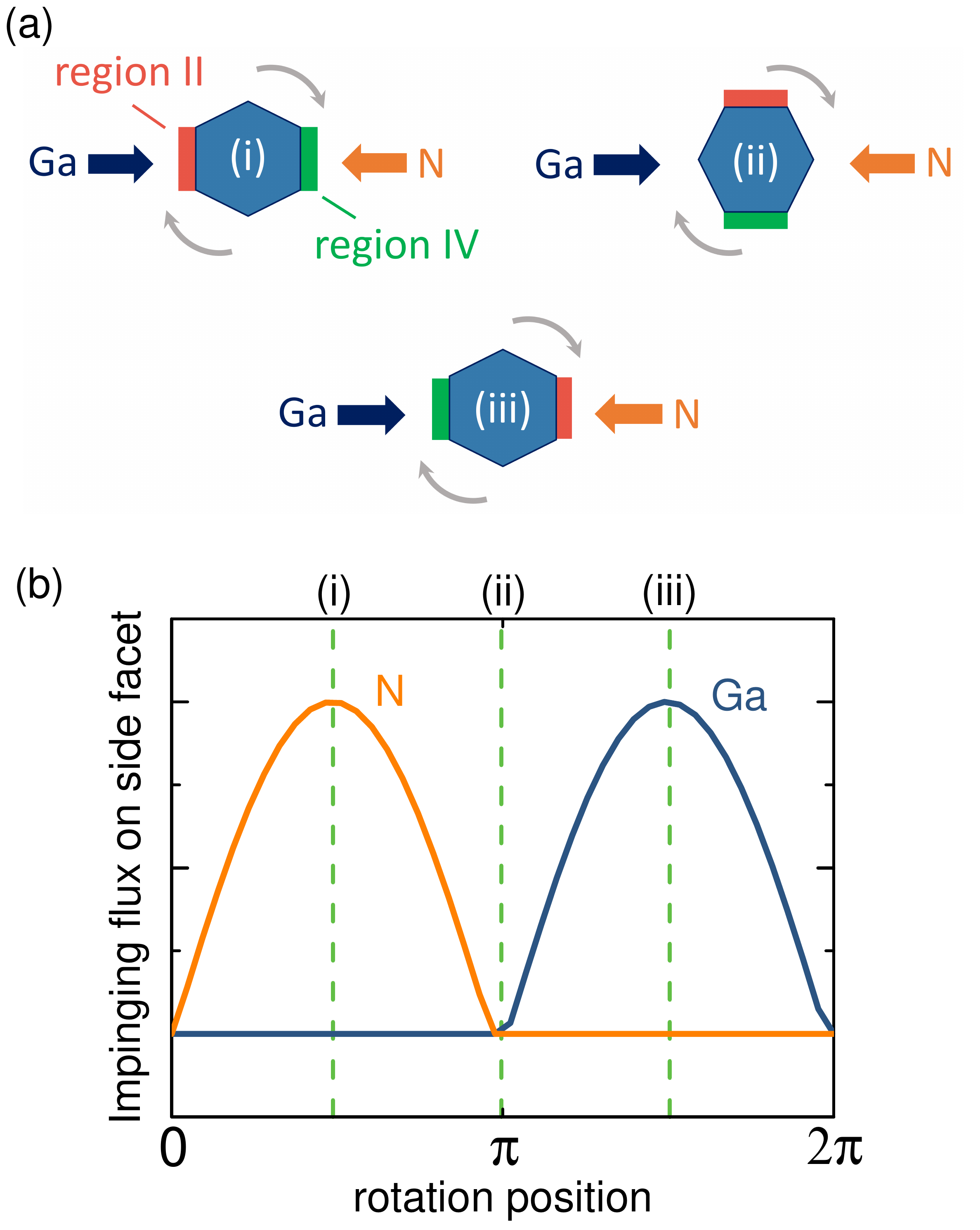}
\caption{(a) The plan view sketches show the orientation of region II and IV with respect to the Ga and N source for three different rotation positions: (i) $\pi/2$, (ii) $\pi$, and (iii) $3\pi/2$ in the case of $\beta=$\;180\grad. (b) The graph shows the Ga and N flux impinging on region IV depending on the rotation position. The dashed green lines indicate the different positions (i)\,--\,(iii) as sketched above.}
\label{fig5}
\end{figure}

In order to simulate substrate rotation in our model, region II and IV have to be rotated with respect to the Ga and N source. The sketches in Fig.\,\ref{fig5}\,(a) show three different rotation positions for the scenario where region II and IV are located on opposite sides ($\beta=$\;180\grad).  
At position (i), region II and region IV are directly exposed to Ga and N, respectively. 
Once rotation is started, the regions move away from the direct beams of the sources. For instance, region IV first rotates out of the N beam until it is neither exposed to N nor to Ga (ii), then rotates into the Ga beam (iii) and subsequently back to the position where only N impinges on the facet. The flux sequence impinging on region IV during one rotation is shown Fig.\,\ref{fig5}\,(b). The dashed green lines indicate the different positions (i)\,--\,(iii) sketched in Fig.\,\ref{fig5}\,(a). 

To implement the continuous rotation in our model, the diffusion equations of region II and IV as they were defined in section \ref{sec:A} have to be modified. Since in the rotation scenario, during one rotation both regions are exposed to Ga and N, the diffusion equations of region II and IV have to include both the incorporation term ${n}/{\tau}$ and the flux term $\chi_\perp J_\text{Ga}$, which are modulated according to the rotation position by time-dependent factors $p$ and $q$, respectively. The modified equations (II) and (IV) are given as

 \renewcommand{\labelenumi}{(\Roman{enumi})}
\begin{enumerate}
\addtocounter{enumi}{1}
\item $\begin{aligned}[t]
     \frac{\partial n_\text{II}(x_\text{II},t)}{\partial t} &= D_\text{side} \frac{\partial^2 n_\text{II}(x_\text{II},t)}{\partial x_\text{II}^2} \\
     &+  p_\text{II}(t)\frac{n_\text{II}(x_\text{II},t)}{\tau_\text{side}}  + q_\text{II}(t)\chi_\perp J_\text{Ga} 
\end{aligned}$
\addtocounter{enumi}{1}
\item $\begin{aligned}[t]
    \frac{\partial n_\text{IV}(x_\text{IV},t)}{\partial t} &= D_\text{side} \frac{\partial^2 n_\text{IV}(x_\text{IV},t)}{\partial x_\text{IV}^2}  \\
    &+ p_\text{IV}(t)\frac{n_\text{IV}(x_\text{IV},t)}{\tau_\text{side}} + q_\text{IV}(t)\chi_\perp J_\text{Ga}
\end{aligned}$
\end{enumerate}
with the time-dependent modulation factors for region II
\begin{equation}
\label{eq:growth rate}
\begin{aligned}
p_\text{II}(t) &= \begin{cases}
\sin(2\pi \nu t -\beta) &\text{for $\sin(2\pi \nu t -\beta)\geq 0$}\\
0 &\text{for $\sin(2\pi \nu t -\beta) < 0$}
\end{cases} \\
q_\text{II}(t) &= \begin{cases}
\sin(2\pi \nu t) & ~~~~~~~\text{for $\sin(2\pi \nu t)\geq 0$}\\
0  & ~~~~~~~\text{for $\sin(2\pi \nu t) < 0$}
\end{cases} 
\end{aligned}
\end{equation}
and region IV
\begin{equation}
\label{eq:growth rate}
\begin{aligned}
p_\text{IV}(t) &= \begin{cases}
\sin(2\pi \nu t) &~~~~~~~\text{for $\sin(2\pi \nu t)\geq 0$}\\
0 &~~~~~~~\text{for $\sin(2\pi \nu t ) < 0$}
\end{cases} \\
q_\text{IV}(t) &= \begin{cases}
\sin(2\pi \nu t+\beta) &\text{for $\sin(2\pi \nu t+\beta)\geq 0$}\\
0 &\text{for $\sin(2\pi \nu t+\beta) < 0$}
\end{cases} 
\end{aligned}
\end{equation}
where $\nu$ is the rotation speed in rotations per minute. The diffusion equations of region I, II, and V are the same as defined in section \ref{sec:A}.
It should be noted that the modified model presented here in section \ref{sec:B} is valid for all azimuthal angles $\beta$ between the N and the Ga source.\\

\begin{figure}[]
\includegraphics[width=0.99\columnwidth, trim= 0 0 0 0]{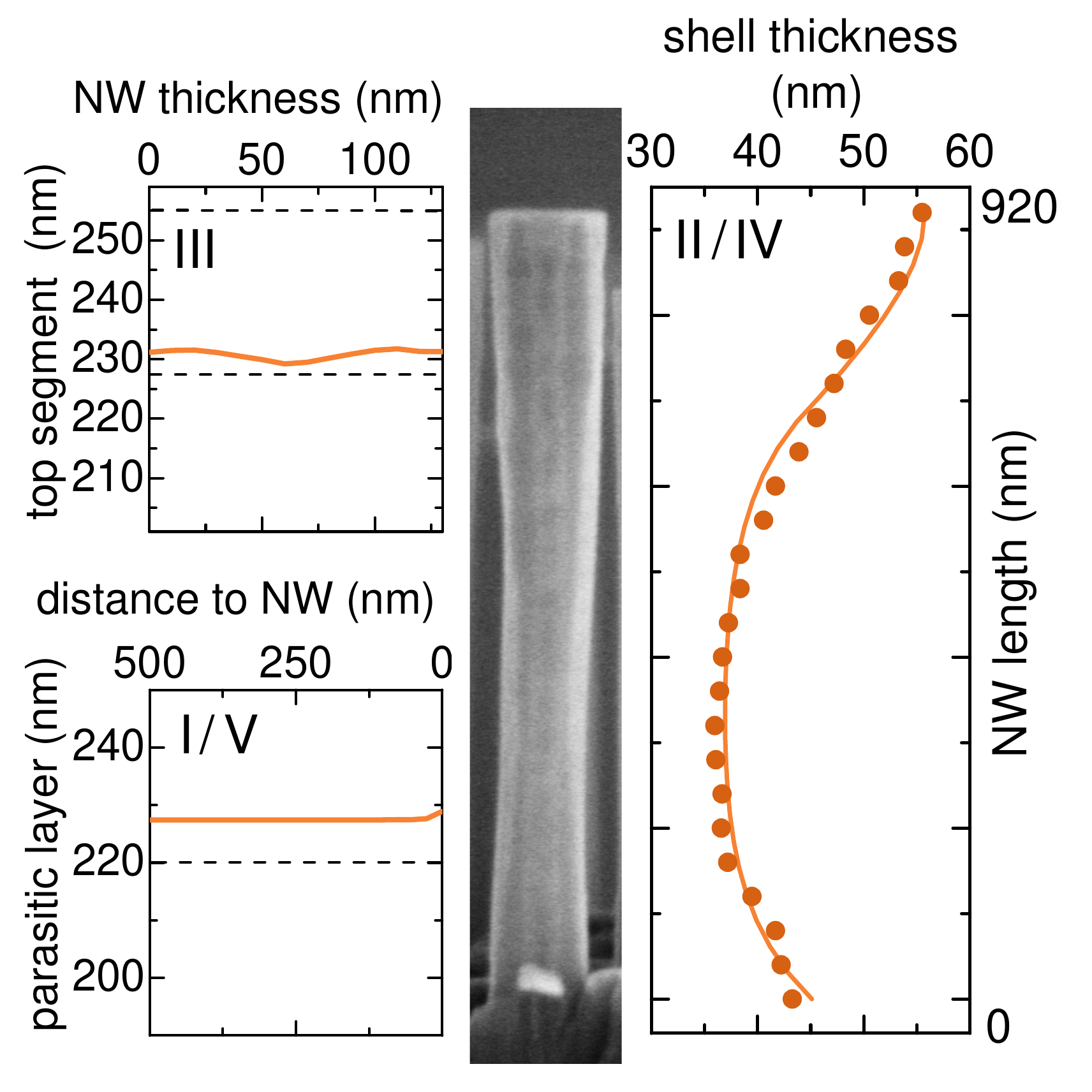}
\caption{The different graphs show the modeled thickness (lines) of a representative NW of the rotated sample B on the side facets (region II and IV), the top facet (region III), and the substrate (region I and V) according to Eq.\,(\ref{eq_GM}). The micrograph in the center depicts the modeled NW. The dots in the diagram of region II/IV indicate the experimental data. The dashed lines in the graph of region III mark the range in which a set of modeling parameters is considered reasonable. The upper dashed line at 255.5\,nm corresponds to the thickness obtained for N-limited growth, the bottom line at 227.5\,nm corresponds to the value obtained for Ga-limited growth. The dashed line in the graph of region I/V with a value of 220\,nm shows the average value found analyzing the thickness of the parasitic layer in areas with low NW density.}
\label{fig6}
\end{figure}

\begin{table}[]
 \caption{Fitting parameters for sample B}
 \label{tab2}
\begin{ruledtabular}
  \begin{tabular}{c c c c c}
sample 	& region  		& $D$ (cm$^2$/s) 		& $\tau$ (s) 	& $\lambda$ (nm) 	\\
\hline
B		& top (III) 	& 3$\times$10$^{-12}$ 	& 0.5 			& 11   				\\
		& side (II+IV) 	& 4$\times$10$^{-10}$ 	& 0.55 			& 155  				\\
		& sub (I+V) 	& 3$\times$10$^{-12}$ 	& 0.5 			& 12 				\\		
\end{tabular}
\end{ruledtabular}
\end{table}

Fig.\,\ref{fig6} shows the modeling of the grown material of a representative NW of the rotated sample B on the side facets (region II and IV), the top facet (region III), and the substrate (region I and V) according to Eq.\,(\ref{eq_GM}).
The angles $\alpha$ and $\beta$ were chosen according to the geometry of our MBE system as explained in the experimental section. 
One can see that the shell thickness of sample B (dots) can be very well described for fitting parameters similar to the ones obtained for the non-rotated case (see Table\,\ref{tab2}).
In general, it should be noted that the two scenarios, with and without rotation, do not necessarily have to have the same diffusion parameters. For instance, in the static non-rotated case of sample E, the Ga adatom concentration on the Ga exposed side facet is higher than for sample B, which results in a higher Ga adatom concentration on the top facet and hence affects the growth rate $n$/$\tau$.
Besides the modeled shell profile for region II and IV, Fig.\,\ref{fig6} depicts the simulated thickness of the parasitic layer (region I/V) and the top segment (region III). 
Since in our model the shadowing of the surface is not taken into account, we use 227.5\,nm as reference value for the layer thickness far away from the NW (see section\,\ref{sec:A}). In the model we obtained a thickness of 227\,nm for distances larger than about 50\,nm from the NW for a set of parameters which resulted in the best description of the shell on the side facet. This value compares well with the experimentally found thickness of the parasitic layer of 220\,nm (dashed line) in areas with low NW density.
Analyzing the thickness of the top segment with an average value of about 231\,nm, one finds that the segment becomes slightly thicker towards the edge of the NW. The higher growth rate in the lateral area of the top facet could be explained by the locally higher Ga adatom concentration induced by adatom diffusion from the Ga exposed side facet, similar to the non-rotated case shown in Fig.\,\ref{fig4}.

\begin{figure}[]
\includegraphics[width=1\columnwidth]{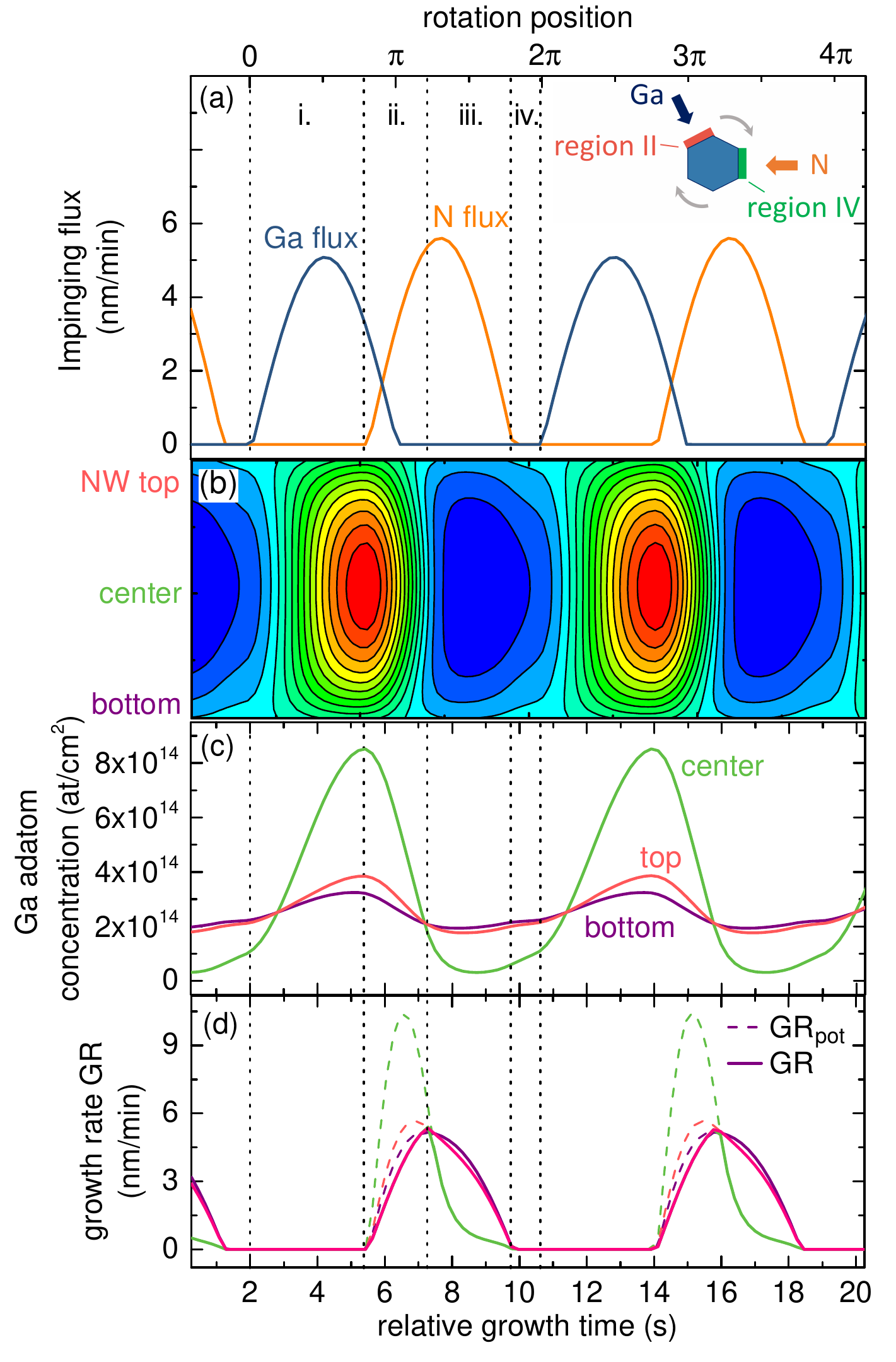}
\caption{(a) Impinging Ga and N fluxes on the side facet region IV for different growth times/rotation positions and for the conditions of sample B. The inset depicts the arrangement of the material sources and the regions II and IV for $\beta= 144^{\circ}$, corresponding to the geometry of our MBE system. (b) The contour plot depicts the modeled Ga adatom concentration of region IV of sample B for different growth times/rotation positions along the whole NW length. The values for the maximum (red) and the minimum (blue) are 8$\times$10$^{14}$\,at/cm$^2$ (0.65 ML) and 3$\times$10$^{13}$\,at/cm$^2$ (0.025 ML), respectively, where the contour lines mark steps of 5.5$\times$10$^{13}$\,at/cm$^2$ (0.045 ML). (c) Detailed evolution of the Ga adatom concentration at the top (red), the center (green), and the bottom (violet) of the NW side facet. The data was extracted from the contour plot data of figure (b). (d) Actual growth rate $GR$ (line) and potential growth rate $GR_\text{pot}$ (dashes) at the top, the center and the bottom of the side facet region IV modeled for sample B. The five vertical dotted lines between 2 and 10.6 seconds in Figs. (a)--(d) indicate the four different rotation phases i.--\,iv. as discussed in the text below.}
\label{fig7}
\end{figure}

In order to better understand the adatom kinetics for the rotation case, in Fig.\,\ref{fig7} we analyze the adatom concentration on the side facets during rotation. It should be mentioned that in our model, the growth on region II and region IV is symmetric under rotation. For simplicity, we focus on region IV in the following discussion.
Fig.\,\ref{fig7}\,(a) shows the Ga and N fluxes impinging on region IV with proceeding growth time, where the maximum values are 5 and 5.6\,nm/min, respectively. The inset depicts the arrangement of the material sources and the regions II and IV for $\beta= 144^{\circ}$, corresponding to the geometry of our MBE system.
The contour plot of Fig.\,\ref{fig7}\,(b) depicts the adatom concentration of region IV for different growth times/rotation positions along the whole NW length, where red is the maximum and blue is the minimum adatom concentration with 8$\times$10$^{14}$\,at/cm$^2$ (0.65 ML) and 3$\times$10$^{13}$\,at/cm$^2$ (0.025 ML), respectively. For a better understanding, Fig.\,\ref{fig7}\,(c) shows the precise evolution of the Ga adatom concentration plotted in Fig.\,\ref{fig7}\,(b) at three different positions on the NW: the top (red), the center (green) and the bottom (violet) of the NW. At the time when region IV is exposed to Ga only, the adatom concentration rises fast in the center and slower at the edges. Subsequently, once region IV rotates away from the Ga beam and into the N beam, the adatom concentration is reduced, fast in the center and slower at the edges, until it is almost homogeneous along the whole NW. 
From this point on the situation on the side facet changes. While decreasing, the adatom concentration is now higher at the top and the bottom edges of the NW and has its minimum in the center. The minimum is reached shortly after the impinging N flux reached its highest value. Subsequently, once region IV rotates into the Ga beam again, the adatom concentration rises and a new rotation cycle begins. 
Fig.\,\ref{fig7}\,(d) shows the actual growth rate $GR$ (line) and the potential growth rate $GR_\text{pot}$ (dashes) at the top, the center and the bottom of the NW side facet region IV. The actual growth rate is defined as $GR={n(x,t)}/{\tau}$ and reaches its maximum value once ${n(x,t)}/{\tau}=\chi J_N$, as introduced in section\,\ref{sec:A}. 
The potential growth rate $GR_\text{pot}$ is not limited by the latter condition and it is assumed that all Ga adatoms can be incorporated, hence, $\chi J_N > {n(x,t)}/{\tau}$ for all times. Or in other words, $GR_\text{pot}$ being larger than $GR$ indicates metal-rich conditions on the facet.
Fig.\,\ref{fig7}\,(d) depicts that once the Ga-wetted surface is exposed to N, both $GR$ as well as $GR_\text{pot}$ increase, where in the beginning $GR_\text{pot}$ is much larger in the center. At the top and the bottom there is not a big difference between the two growth rates. Once region IV passes the N peak flux and rotates away from the N source, $GR$ and $GR_\text{pot}$ are identical.\\

In general, during one rotation period, a side facet passes through four different phases.
The first phase is the wetting phase from 2 to 5.3\,s (0\,--\,3/4$\pi$), where Ga is deposited on a side facet which is still shadowed from the N beam. While the facet rotates away from the Ga source, Ga adatoms diffuse towards the NW top and the substrate due to gradients in the chemical potential. These diffusion processes cause the belly shape of the adatom concentration along the NW, similar to the situation on region II for the non-rotated case in section\,\ref{sec:A}.
The second phase is the metal-rich growth phase between 5.3 and 7.2\,s (3/4$\pi$\,--\,6/5$\pi$). 
Once the facet rotates into the N beam, growth takes place, as shown in Fig.\,\ref{fig7}\,(d). However, in the beginning the growth rate is still low due to the large angle between the facet and the direct N beam and there are more Ga adatoms on the surface than N adatoms being provided by the impinging N flux. The consequence are Ga-rich growth conditions where the growth rate $GR$ is limited by the impinging N flux and is the same along the whole NW length. The Ga excess during that rotation phase is illustrated in Fig.\,\ref{fig7}\,(d) where $GR_\text{pot}$ is higher than $GR$. The Ga adatoms which are not incorporated are free to diffuse away, which leads to a flattening of the belly shape. 

Only with further rotation away from the Ga source and into the N beam, N-rich conditions are established on the side facet. This is the third phase, the N-rich growth phase from 7.2 to 9.8\,s (6/5$\pi$\,--\,9/5$\pi$). The priorly deposited Ga is consumed and the chemical potential on the side facet is lowered, which leads to diffusion from the top facet and the substrate onto the side facet.
In this phase where $\chi J_N > {n(x,t)}/{\tau}$, the growth rate is higher at the top and the bottom of the side facet resulting in the experimentally observed hourglass shape of the shell.
The fourth and last phase is the dissociation phase between 9.8 and 10.6\,s (9/5$\pi$\,--\,2$\pi$). With ongoing rotation, depending on the arrangement of the sources, there may be a rotation phase during which the side facet is neither exposed to Ga nor to N. In this phase, the probability for thermal dissociation of the grown material is the highest. However, in the case of GaN and for substrate temperatures far below 800\grad, the dissociation rate is negligibly small.

\begin{figure}[]
\includegraphics[width=1\columnwidth, trim= 0 0 0 0]{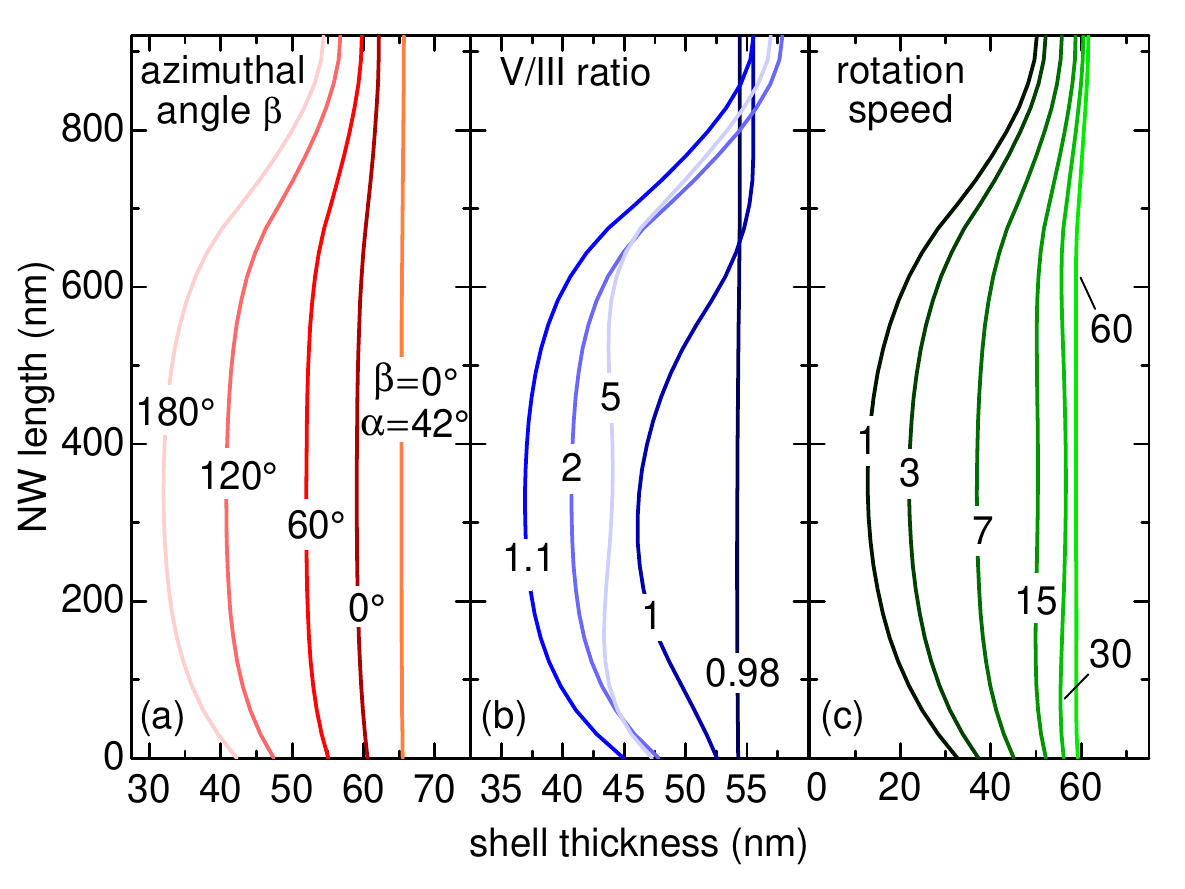}
\caption{Modeled evolution of the shell thickness along the NW for (a) different angles $\beta$ between the Ga and N source, (b) different V/III ratios, and (c) different rotation speeds. If not indicated otherwise in the graph, the modeling is based on the growth and fitting parameters of sample B as mentioned in the experimental section and table\,\ref{tab2}, respectively. The V/III ratio was varied by changing the N flux and maintaining a constant Ga flux of $\Phi_\text{Ga}=8.2~\text{nm/min}$.}
\label{fig8}
\end{figure}

The analysis in Fig.\,\ref{fig7} shows that the Ga adatom kinetics on the different NW facets is very complex during a common growth with substrate rotation and that it has a major influence on the final NW morphology. Also for the case with substrate rotation, we conclude that the main driving force for the pronounced adatom diffusion are large differences in the chemical potential on the various facets originating from the different azimuthal angles of the material sources.
The experimental results shown in Fig.\,\ref{fig1} and \ref{fig2} as well as our conclusions from the modeling of sample B and E reveal that it seems not to be straightforward to experimentally obtain homogeneous shell growth.
However, since the model describes very well the experimentally observed shell profile as well as the growth on top facet and substrate, it allows to further explore different growth conditions which were experimentally not accessible due to technical restrictions. In Fig.\,\ref{fig8} we analyze the evolution of the shell shape for different azimuthal angles, V/III ratios and rotation speeds at otherwise similar parameters as obtained for the modeling of sample B (see table\,\ref{tab2}). 
Since the growth in region II and IV is identical, also for the analysis in Fig.\,\ref{fig8} we focus on region IV.\\

Fig.\,\ref{fig8}\,(a) shows the evolution of the shell morphology for different angles $\beta$ between the Ga and N source. 
For a large $\beta$ of 180\,-\,120\grad the shell depicts a distinct hourglass shape, similar to the morphology of sample B. For angles of 60\grad and smaller, the hourglass profile of the shell becomes less pronounced, however, the shell thickness still increases visibly towards the NW top and the bottom. The most homogeneous shell is obtained for a $\beta$ of 0\grad, where the Ga and N beam impinge from the same direction onto the side facet. For this source configuration, the differences in chemical potentials of the regions III, IV, and V are very small, leading to only little Ga adatom diffusion between the different facets. Hence, most of the material grows where it impinges.
For the fitting parameters of sample B, a perfectly smooth shell with the same thickness along the whole NW length is only achieved for a $\beta$ of 0\grad and a polar angle $\alpha$ of 42\grad (instead of 37.5\grad like in our MBE system). 
In general, if the incorporation/growth rates $n/\tau$ were the same in all regions, an equilibrium of the chemical potentials would be achieved for an angle $\alpha$ of 45\grad. For this angle, the impinging material fluxes impinging on the top facet, the substrate, and the respective side facets are identical. Since according to the fitting parameters obtained for sample B, the growth rate is slightly smaller on the side facet due to a higher adatom lifetime $\tau_\text{side}$, an equality of chemical potentials is achieved for $\alpha=$\,42\grad.

Fig.\,\ref{fig8}\,(a) clarifies that, indeed, the shell morphology strongly depends on the azimuthal angles between the different sources and rather homogeneous shell morphologies can be achieved once the Ga and N sources are arranged next to each other.     
However, arranging the material sources close to each other is not necessarily realizable in standard MBE systems, where the different source ports are usually separated by an azimuthal angle of at least 30\grad. Especially, once numerous material sources are needed for the growth of more complex and/or doped shell structures, it might be challenging to optimize the source configuration. 
This drawback raises the question whether a more homogeneous shell morphology can also be achieved by simply optimizing e.\,g.~the V/III ratio or the rotation speed instead of physically rearranging the material sources.

Fig.\,\ref{fig8}\,(b) depicts the shell profiles for V/III ratios from 0.98 to 5, hence, moving from nominally Ga-rich to N-rich conditions. For the modeling, the Ga flux is kept constant at $\Phi_\text{Ga}=8.2~\text{nm/min}$. 
For V/III ratios of 1 and above, the NW exhibits a very pronounced hour glass shape. For the highest V/III ratio of 5, the shell shows in addition a belly in the NW center. For a ratio of 1, the top part of the shell is rather straight at a value of about 55\,nm. For even lower V/III ratios, the hour glass shape diminishes until a homogeneous shell thickness of 54\,nm along the whole NW is achieved for a V/III ratio of 0.98.
The belly shape in the center of the shell for high V/III ratios can be explained by the fact that with increasing N flux, the metal-rich growth phase on the side facet becomes much shorter. Hence, the belly shape in the Ga adatom concentration that developed during the wetting phase is preserved until the early stage of the N-rich growth phase and manifests itself in the shell morphology. Only with further N-rich growth, the Ga atom concentration on the side facet is lowered and the Ga diffusion from the NW top and the substrate onto the side facet increases, resulting in an increased shell thickness towards the NW top and the bottom.
For lower V/III ratios, the metal-rich growth phase is much longer and most of the initially deposited Ga has diffused away until N-rich conditions are reached and a pronounced hourglass shape develops as discussed for Fig.\,\ref{fig4}.

The strong increase in the shell thickness between V/III ratios of 1.1 and 0.98 results from an increased diffusion of excess Ga onto the side facet once metal-rich growth conditions are reached on the top facet and/or the substrate. For a V/III ratio of 1, e.\,g. the top part of the shell is grown under N-limited conditions due to the diffusion of excess Ga from the Ga exposed facet towards the top facet and from the top facet onto the side facet where growth takes place.
This increase in shell thickness was also experimentally observed comparing sample A and B. Once the V/III ratio is lower than 1, metal-rich conditions are reached on all facets at any time of the rotation resulting in a very homogeneous shell. However, at V/III ratios <\,1, wide, trapezoid-shaped top segments form, as seen for sample A and reported before \cite{Songmuang2010}. Hence, as already observed experimentally for the samples A\,--\,C in Fig.\,\ref{fig1}, it does not seem possible to achieve a straight shell morphology without enlarged top segments by optimizing the V/III ratio.\par

The last important parameter we want to address is the rotation speed. 
Fig.\,\ref{fig8}\,(c) shows that with increasing rotation speed from 1 to 15\,rpm the pronounced concave profile of the shell vanishes and the shell becomes more homogeneous along the NW, however, the shell thickness still increases significantly towards the NW top. For rotation speeds higher than 15\,rpm the shell develops a slight belly in the NW center. Only for rotation speeds of 60\,rpm and higher a shell with a rather homogeneous thickness along the whole NW length is achieved.
In general, the shell thickness increases with increasing rotation speed. 
As discussed for Fig.\,\ref{fig7}, the hourglass shape of the shell found for low rotation speeds
can be explained by the fact that the majority of the Ga that contributes to the growth diffuses from the top and the bottom onto the side facet, while most of the directly deposited Ga has already diffused away. However, the faster the rotation, the less of the directly deposited Ga is able to diffuse away during the wetting and metal-rich phase.  
Hence, once N-rich conditions are reached, the adatom concentration on the side facet is still rather high and homogeneous along the NW leading to an increased and rather uniform shell thickness.
Moreover, the more Ga remains on the side facet until N-rich conditions are established, the lower is the gradient in chemical potential between side facet and the other facets, which results in only little Ga diffusion from the NW top and the substrate during the N-rich phase. This explains the comparatively small increase in shell thickness towards the NW top and the NW bottom for high rotation speeds.

The analysis of Fig.\,\ref{fig8}\,(c) shows that a homogeneous shell morphology can also be achieved by high rotation speeds >\,60\,rpm. It should be noted that such high rotation speeds are not realizable in all MBE systems. In our case, for example, due to restrictions of the driving motor, we were only able to experimentally test rotation speeds up to 13\,rpm, where we still observed a pronounced hourglass shape of the shell. \\

\section{SUMMARY AND CONCLUSIONS}\label{sec:summary}

In this combined experimental and theoretical study, we have investigated in detail the effect of the sequential deposition in MBE on the growth of GaN shells around GaN NWs. Our experimental results showed that it is not straightforward to obtain shells with homogeneous thickness along the whole NW length. 

By modeling the shell growth \textit{with} and \textit{without} substrate rotation, we found that the hourglass shape observed for most growth conditions is the result of strong Ga adatom diffusion processes on the NW. These diffusion processes are decisively influenced by the fact that 
growth is possible only under directly impinging N, while Ga adatoms diffuse between the sidewalls and the top facet as well as the substrate, but not between adjacent sidewall facets. 
Modeling shell growth for various conditions, we found that homogeneous shells can be achieved for small azimuthal angles between Ga and N, where both materials impinge from the same direction.
However, the positioning of material sources next to each other is not always practical or feasible, especially considering the growth of complex core-shell structures where ternary or quaternary alloys and doping are involved.
As a solution to this impediment, we showed that also for very high rotation speeds rather homogeneous shells can be obtained. However, the necessary rotation speeds are rather fast and might not be realizable in every MBE system.

Beyond the pragmatic benefit of providing guidance for the design of growth protocols resulting in homogeneous shells, our model has also enabled a comprehensive understanding of the adatom concentration on the NW at all times during the growth. 
We found that the growth on the side facets can be categorized in four different phases: the wetting phase, the metal-rich growth phase, the N-rich growth phase, and the dissociation phase.
The absolute and relative length of these phases during one rotation period may not only affect the thickness homogeneity of the shell along the NW as discussed above, but also its local surface roughness.
The fact that for larger azimuthal angles, Ga and N are deposited at different times during the rotation leads to an extensive Ga diffusion during the wetting and the metal-rich growth phase, similar to an MEE growth process \cite{Foxon2009,Horikoshi1999a}. This may explain the smooth side facets generally observed for NWs also for very low substrate temperatures and nominally N-rich growth conditions.
The striking difference to growth processes on planar samples is that in our case, diffusion takes place between different regions, i.\,e.~the sidewall vs. the top facet and substrate, out of which on one N impinges not continuously, resulting in complex gradients in chemical potential that are modulated in time by substrate rotation. 

Our analysis has shown that even for a binary material like GaN, shell growth by MBE is very complex. This finding raises the question how the shell growth of ternary or even more complex alloys may be influenced by the chamber geometry. For example, in the case of (In,Ga)N, where Ga is incorporated preferentially over In and In segregation may occur, the In incorporation might strongly be affected by the different characteristics of the four rotation phases. 

In general, the strength of our model is that it not only considers all relevant regions like the different NW facets and the substrate, but also the full MBE geometry by taking into account the different time-dependent orientations of the side facets with respect to the material sources. This comprehensiveness of the model provides a deep understanding of diffusion processes and the resulting adatom concentration, and could be applied to other 3D structures and material systems.

\section*{ACKNOWLEDGMENT}
We are grateful to C.~Stemmler, K.~Morgenroth and M.~Höricke for the maintenance of the MBE system, and A.-K. Bluhm for the SE micrographs. Furthermore, we are thankful to V.~Kaganer for a critical reading of the manuscript as well as to O.~Brandt and T. Auzelle for fruitful discussions. Sergio Fernández-Garrido acknowledges the partial financial support received through the Spanish program Ramón y Cajal (co-financed by the European Social Fund) under grant RYC-2016-19509 from Ministerio de Ciencia, Innovación y Universidades.

\bibliography{simulation}

\end{document}